\documentclass[%
 reprint,
 amsmath,amssymb,
 aps,
]{revtex4-2}
\usepackage{float}
\usepackage{graphicx}
\usepackage{dcolumn}
\usepackage{bm}
\usepackage{hyperref}

\begin{document}

\preprint{APS/123-QED}

\title{Double Copy Map for Double Field Theory on an Arbitrary Constant Background}

\author{Rasim Yılmaz}%
\email{ryilmaz@metu.edu.tr}
\affiliation{%
Department of Physics, Faculty of Arts and Sciences\\
Middle East Technical University, 06800, Ankara, Türkiye
}%

\date{\today}

\begin{abstract}
Double field theory (DFT) can be constructed up to third order as the double copy of Yang-Mills theory. In this construction, one starts with Yang-Mills theory on Minkowski space and the resulting DFT action is also defined on Minkowski space. In this work, I extend this double copy procedure to obtain a DFT action on a constant DFT background starting from two Yang-Mills actions constructed around nontrivial background gauge fields on Minkowski space.

\end{abstract}

\maketitle

\section{Introduction}
The Bern-Carrasco-Johansson double copy framework \cite{bcj, Bern2010} is a fascinating and powerful idea that reveals deep connections between gravity and gauge theories at the level of scattering amplitudes. This idea takes its origin from the KLT relations \cite{klt}, which state that closed string tree-level amplitudes can be obtained from open string amplitudes. In addition to being at the center of the modern amplitude program (\cite{bcjreview1}, \cite{bcjreview2}), this framework also provides new insights about the relation of gravity and gauge theories. In recent years, the double copy idea has also been extended to the level of classical solutions (see, for example, \cite{Monteiro, Luna, Carrillo, Kim2019}). Usually, to relate classical solutions of gravity to the classical solutions of Yang-Mills theory, one starts with a perturbative metric around a Minkowski space background and relates the perturbation part of the metric to the two copies of Yang-Mills gauge fields. However, in \cite{Adamo} and \cite{abbas}, the authors show that it is also possible to construct a perturbative gravity theory around a nontrivial background metric by a double copy procedure. There are two approaches that one can use for this construction, which are called the type-A and the type-B double copies by the authors of \cite{abbas}. In the type-B double copy one starts with the Yang-Mills field around a non-dynamical curved metric and constructs the gravity theory around the same curved metric. In the type-A double copy, one starts with a Yang-Mills field around a non-trivial background gauge field in Minkowski space and obtains a gravity theory around a non-dynamical curved metric. In this procedure, the nontrivial background gauge field creates the non-dynamical curved background metric by the double copy procedure and the dynamical gauge field creates the perturbative part of the metric for the gravity theory. \\

In recent years, the applications of double copy idea have been extended to many different theories (see \cite{Berman:2020xvs, Angus:2021zhy, Cho:2021nim, Berman, Jonke}). There are also some applications of the double copy procedure at the level of actions. One of the interesting results on this subject relates the usual Yang-Mills action on a Minkowski space to the DFT action on a Minkowski space \cite{jaram}. DFT is a modern framework in theoretical physics to implement T-duality, a symmetry of string theory, to the field theories defined on a double configuration space \cite{hulldft, backindepdft, Hohm:2010pp}. Double configuration space includes the dual coordinates in addition to the usual coordinates and doubles the number of dimensions to have a manifest $O(D,D)$ duality (see \cite{aldaz} for a detailed review). The double structure of DFT implies a relation with the double copy idea and there are some works based on this relation, \cite{Lee,cho,lescanohetero}. Recently the idea to construct DFT as double copy of Yang-Mills is also extended to the higher derivative theories (see \cite{Lescano:2023pai, Lescano:2024gma, Lescano:2024lwn, rasim}), and to the heterotic sector of DFT (see \cite{cho, lescanohetero, rasim2}). There are also algebraic approaches to relate DFT action to the double copy of Yang-Mills theory (see for example \cite{Bonezzi,Bonezzi2, Bonezzi3, Bonezzi4}). \\

In \cite{jaram}, the authors take the Yang-Mills action on a Minkowski space as a starting point and the DFT action they obtain is defined on the Minkowski space. The purpose of this work is to extend this analysis to the DFT action constructed around an arbitrary constant DFT background. In section 2, I review the type-A double copy idea for classical solutions and in section 3, I attempt to apply this idea in the context of DFT. The results of this section imply that a naive application of double copy procedure to a Yang-Mills action constructed around an arbitrary constant gauge field and Minkowski metric cannot give a DFT action around arbitrary constant DFT background. The main result of this work is given in section 4, where I follow a slightly different route from the usual literature on double copy construction of DFT. Although this different route can be interpreted as a slight notation change, it seems to give a more natural way to implement double copy idea for a DFT action on an arbitrary constant background. In that section, I start with two Yang-Mills actions which are perturbatively constructed around nontrivial background gauge fields, on Minkowski space and obtain DFT action defined on an arbitrary constant DFT background by using double copy procedure. The last section is devoted to the discussions of the results, their importance for the current literature, and possible implications for the future works.
\section{Type-A Double Copy}
Here, I review the type-A double copy idea, following \cite{abbas}. In the usual literature, the starting point of the double copy for classical solutions is the Kerr-Schild metric. Consider the full metric as,
\begin{equation}
    g_{\mu \nu}=\bar{g}_{\mu \nu}+ \kappa h_{\mu \nu},
\label{fulmetric}
\end{equation}
where the perturbation part is given by,
\begin{equation}
    h_{\mu \nu}=\frac{\kappa}{2} \phi k_\mu k_\nu,
\label{graviton}
\end{equation}
with $\phi(x)$ is scalar function, $k_\mu$ is a 4-vector that is null and geodesic both with respect to the background and the full metric \cite{abbas}. So one has,
\begin{equation}
    g^{\mu \nu} k_\mu k_\nu=\bar{g}^{\mu \nu} k_\mu k_\nu=0, \quad k^\mu D_\mu k_\nu=0,
\end{equation}
where $D_\mu$ is covariant derivative associated with the background metric. Choosing the background metric as Minkowski, $\bar{g}_{\mu \nu}=\eta_{\mu \nu}$, one finds a linear form for the Ricci tensor of the full metric (\ref{fulmetric}) as,
\begin{equation}
    R_\nu^\mu = \frac{\kappa}{2} \partial_\rho\left(\partial_\nu h^{\mu \rho}+\partial^\mu h_\nu^\rho-\partial^\rho h_\nu^\mu\right).
\end{equation}
For a given scalar field $\phi$ and Kerr-Schild vector $k_\mu$, a Yang-Mills gauge field can be defined as single copy of (\ref{graviton}) as \cite{abbas},
\begin{equation}
    A_\mu^a=c^a \phi k_\mu,
\label{singlecopy}
\end{equation}
where $c^a$ is an arbitrary constant color vector. The upshot is that when $g_{\mu \nu} = \eta_{\mu \nu}+\kappa h_{\mu \nu}$ with $h_{\mu \nu}$ defined as (\ref{graviton}), the corresponding single copy (\ref{singlecopy}) satisfies the linearised Yang-Mills equations, namely,
\begin{equation}
    \partial^\mu F_{\mu \nu}^a=0.
\end{equation}
This double copy construction is introduced for the Kerr-Schild metric which is defined around a Minkowski background. As shown in \cite{abbas}, this construction can also be extended to the curved background. There are two possible approaches in \cite{abbas}, namely, type-A and type-B double copies. Here, I concentrate on type-A double copy. In this approach, a gauge field has a non-trivial background field $\bar{A}^a_\mu$ in Minkowski space, and the corresponding double copy gives a perturbative metric defined on a non-dynamical curved background $\bar{g}_{\mu \nu}$, where $\bar{g}_{\mu \nu}$ and $\bar{A}^a_\mu$ are also related by a double copy relation \cite{abbas}. The idea can be schematically given as \cite{abbas},
\begin{figure}[H]
    \centering
    \includegraphics[width=0.7\linewidth]{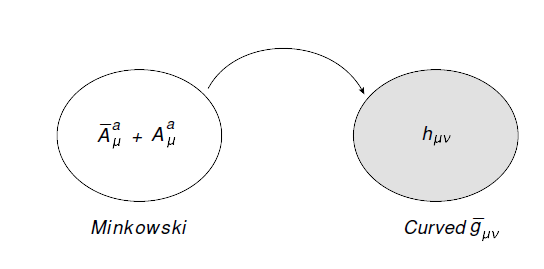}
    \caption{Type-A double copy}
    \label{fig:enter-label}
\end{figure}
A simple example can be constructed, by starting with single Kerr-Schild metrics around Minkowski space. One can split up the solutions according to,
\begin{equation}
    \begin{aligned}
g_{\mu \nu} & =\eta_{\mu \nu}+\phi k_\mu k_\nu \\
& =\eta_{\mu \nu}+\phi_1 k_\mu k_\nu+\phi_2 k_\mu k_\nu,
\end{aligned}
\label{typeametric}
\end{equation}
where,
\begin{equation}
    \phi_1=\xi \phi, \quad \phi_2=(1-\xi) \phi, \quad 0 \leq \xi \leq 1 .
\end{equation}
The corresponding single copy for the gauge field reads,
\begin{equation}
    A_\mu^a=c^a\left[\phi_1 k_\mu+\phi_2 k_\mu\right].
\label{typeagauge}
\end{equation}
One can interpret equation (\ref{typeametric}) as,
\begin{equation}
    g_{\mu \nu}=\bar{g}_{\mu \nu}+\tilde{h}_{\mu \nu},
\end{equation}
where,
\begin{equation}
    \bar{g}_{\mu \nu}=\eta_{\mu \nu}+\phi_1 k_\mu k_\nu \quad \text{and} \quad \tilde{h}_{\mu \nu}=\phi_2 k_\mu k_\nu.
\label{metricsplit}
\end{equation}
Similarly, (\ref{typeagauge}) can be interpreted as,
\begin{equation}
    A_\mu^a=\bar{A}_\mu^a+\tilde{A}_\mu^a,
\end{equation}
where,
\begin{equation}
    \bar{A}_\mu^a=c^a \phi_1 k_\mu \quad \text{and} \quad \tilde{A}_\mu^a=c^a \phi_2 k_\mu.
\label{gaugesplit}
\end{equation}
Hence, (\ref{metricsplit}) and (\ref{gaugesplit}) imply a double copy relation as
\begin{equation}
    \bar{A}_\mu^a \rightarrow \bar{g}_{\mu \nu} \quad \text{and} \quad \tilde{A}_\mu^a \rightarrow \tilde{h}_{\mu \nu}.
\end{equation}
This is an example of type-A curved space double copy for classical solutions. A Yang-Mills gauge field defined around a non-trivial background gauge field is copied to the perturbative metric field with a non-trivial background, where the background fields are also related by double copy \cite{abbas}.
\section{Type-A Double Copy for DFT}
Inspired by the results of the previous section, here, I apply the double copy prescription to Yang-Mills action around a nontrivial background gauge field. I start with the standard Yang-Mills action and then expand the gauge field $A_\mu{}^a$ around a constant background $\bar{A}_\mu{}^a$ as,
\begin{equation}
    \mathcal{A}_\mu{}^a = {A}_\mu{}^a + a_\mu{}^a,
\end{equation}
where $a_\mu{}^a$ is a small perturbation. Then, the field strength tensor becomes
\begin{equation}
    F_{\mu \nu}{ }^a=\partial_\mu a_\nu{ }^a-\partial_\nu a_\mu{ }^a+g_{\mathrm{YM}} f^a{ }_{b c} ({A}_\mu{}^b + a_\mu{}^b)({A}_\nu{}^c + a_\nu{}^c).
\label{fieldstrength}
\end{equation}
By using (\ref{fieldstrength}), Yang-Mills action reads,
\begin{equation}
     S_{\mathrm{YM}}=-\frac{1}{4} \int d^D x \kappa_{a b} (F_{\mu \nu}{ }^a )(F^{ \mu \nu b} )
\end{equation}
which can be expanded as,
\begin{equation}
    S_{\mathrm{YM}} = S^{0}_{\mathrm{YM}} + S^{(1)}_{\mathrm{YM}} + S^{(2)}_{\mathrm{YM}}+ \cdots,
\end{equation}
where $S^{0}_{\mathrm{YM}}$ is the action for the background field, $S^{(1)}_{\mathrm{YM}}$ is linear in $a_\mu{}^a$ and $S^{(2)}_{\mathrm{YM}}$ corresponds to quadratic terms. The quadratic part is of particular interest and is given by,
\begin{equation}
\begin{aligned}
    S^{(2)}_{\mathrm{YM}}=-\frac{1}{4} \int_{x} \bigg ( & \kappa_{a b} (\partial_\mu a_\nu{ }^a-\partial_\nu a_\mu{ }^a)(\partial^\mu a^{\nu b}-\partial^\nu a^{\mu b}) \\
    &+4g_{\mathrm{YM}} f_{a b c} (\partial_\mu a_\nu{ }^a-\partial_\nu a_\mu{ }^a) {A}^{\mu b} {a}^{\nu c} \\
    & +2g^2_{\mathrm{YM}} \kappa^{a b} f_{a c d} f_{b e f} \left({A}_\mu{}^c {A}_\nu{}^d a^{\mu e} a^{\nu f} \right. \\ 
    &\left. + {A}_\mu{}^c a_{\nu}{}^d  {A}^{\mu e} a^{\nu f} + {A}_\mu{}^c a_\nu{}^d a^{\mu e} {A}^{\nu f}\right) \bigg),
\end{aligned}
\label{backymaction}
\end{equation}
where the indices are raised and lowered by Minkowski metric. Now, I apply double copy prescription to (\ref{backymaction}). The corresponding double copy maps are given as \cite{jaram},
\begin{equation}
\begin{aligned}
    & a_\mu{ }^a(k) \rightarrow e_{\mu \nu}(k, \bar{k}), \\
    & \kappa_{a b} \rightarrow \frac{1}{2} \bar{\Pi}^{\nu \sigma}(\bar{k}), \\
    & f_{abc} \rightarrow  \frac{i}{4} \bar{\pi}^{{\nu} {\sigma} {\tau}}\left(\bar{k}_1, \bar{k}_2, \bar{k}_3\right),
\end{aligned}  
\label{dcmap1}
\end{equation}
where the cubic projector is defined by
\begin{equation}
    \pi^{\nu \sigma \tau}\left(k_i\right) \equiv \eta^{\nu \sigma} \bar{k}_{12}^\tau+\eta^{\sigma \tau} \bar{k}_{23}^\nu+\eta^{\tau \nu} \bar{k}_{31}^\sigma.
\label{cubicprojector}
\end{equation}
Apart from these, I also define a double copy map for the background gauge field as
\begin{equation}
    {A}_\mu{ }^a(k) \rightarrow E_{\mu \nu}(k, \bar{k}).
\label{dcmap2}
\end{equation}
Now, I substitute (\ref{dcmap1}) and (\ref{dcmap2}) into (\ref{backymaction}). To apply this prescription, it is convenient to pass to the momentum space. Let me divide the action (\ref{backymaction}) into three and start with the first part, namely,
\begin{equation}
   S^{(2)}_{\mathrm{1YM}}=-\frac{1}{4} \int_{x}  ( \kappa_{a b} (\partial_\mu a_\nu{ }^a-\partial_\nu a_\mu{ }^a)(\partial^\mu a^{\nu b}-\partial^\nu a^{\mu b}) ).
\label{backymaction1}
\end{equation}
The double copy of this action is familiar from \cite{jaram}. Passing to the momentum space, applying the double copy prescription, (\ref{backymaction1}) becomes
\begin{equation}
\begin{aligned}
S_{\mathrm{1DC}}^{(2)}=-\frac{1}{4} \int_{k, \bar{k}} & \left(k^2 e^{\mu \nu} e_{\mu {\nu}}-k^\mu k^\rho e_{\mu {\nu}} e_\rho{}^{{\nu}}-\bar{k}^{{\nu}} \bar{k}^{{\sigma}} e_{\mu {\nu}} e^\mu{ }_{{\sigma}}\right. \\
& \left.-k^2 \phi^{2} +2 \phi k^{\mu} \bar{k}^\nu e_{\mu \bar{\nu}}\right),
\end{aligned}
\label{sdc1}
\end{equation}
in momentum space. Passing to the position space, it reads
\begin{equation}
    \begin{aligned}
S_{\mathrm{1DC}}^{(2)}=  \frac{1}{4}  \int d^D x &d^D  \bar{x}\left(e^{\mu {\nu}} \square e_{\mu {\nu}}+\partial^\mu e_{\mu {\nu}} \partial^\rho e_\rho{}^{{\nu}}\right. \\
& \left.+\bar{\partial}^{{\nu}} e_{\mu {\nu}} \bar{\partial}^{{\sigma}} e^\mu{}_{{\sigma}}-\phi \square \phi+2 \phi \partial^\mu \bar{\partial}^{{\nu}} e_{\mu {\nu}}\right).
\end{aligned}
\end{equation}
Then, I can continue with the second piece of (\ref{backymaction}), namely,
\begin{equation}
     S^{(2)}_{\mathrm{2YM}}=-\int_{x} g_{\mathrm{YM}} f_{a b c} (\partial_\mu a_\rho{ }^a-\partial_\rho a_\mu{ }^a) {A}^{\mu b} a^{\rho c}
\end{equation}
Upon Fourier transforming to momentum space this becomes,
\begin{equation}
   S_{2 \mathrm{YM}}^{(2)}=-i g_{\mathrm{YM}} \int_{k}f_{abc}\left(k_\rho a_\mu^a(k)-k_\mu a_\rho^a(k)\right) {A}^{\mu b} {a}^{\rho c}(-k).
\label{secondmomspaceym}
\end{equation}
Imposing the double copy prescriptions (\ref{dcmap1}),(\ref{dcmap2}) on (\ref{secondmomspaceym}), with $g_{\mathrm{YM}} \rightarrow \frac{1}{2}$, one gets,
\begin{equation}
\begin{aligned}
    S_{2 \mathrm{DC}}^{(2)}=\frac{1}{8} & \bigg( \int_{k,\bar{k}}  \bar{\pi}^{\nu \sigma \kappa} \eta^{\mu \tau}k^\rho e_{\mu \nu}(k,\bar{k})E_{\tau \sigma}e_{\rho \kappa}(-k,-\bar{k})  \\
    & \quad \quad -\bar{\pi}^{\nu \sigma \kappa} \eta^{\rho \tau}k^\mu e_{\mu \nu}(k,\bar{k})E_{\tau \sigma}e_{\rho \kappa}(-k,-\bar{k}) \bigg).
\end{aligned} 
\end{equation}
Here, some problems arise. One can see that this proposal is incomplete since $\bar{\pi}^{\nu \sigma \kappa}$ depends on three momenta, while in the above terms there are only two $e$ fields. Although I tried several ways to continue from this point, I could not find a consistent way to obtain DFT action on an arbitrary constant background following this route. Instead, in the next section, I offer a slightly different way of thinking the double copy procedure in DFT context, and show how it can help us to construct DFT action on an arbitrary constant DFT background.
\section{Type-A Double Copy with an Alternative Formulation}
Now I introduce an alternative point of view for the double copy procedure in DFT context. The key observation is that while in the usual double copy procedure for scattering amplitudes one starts with two (same or different) Yang-Mills theories to obtain gravity amplitudes, the double copy procedure for DFT action uses only one Yang-Mills action and maps its color indices to the second set of indices in DFT. In this section, I start with two Yang-Mills actions and introduce double copy maps to obtain the usual DFT action. Consider two Yang-Mills actions given by,
\begin{equation}
  \begin{aligned}
      S_{\mathrm{YM1}} & =-\int d^D x \left(\frac{1}{4} \kappa_{a b} F^{\mu \rho a} F_{\mu \rho}{ }^b \right), \\
      S_{\mathrm{YM2}} & =- \int d^D \bar{x} \left(\frac{1}{4}\bar{\kappa}_{a b} \bar{F}^{\nu \sigma a} \bar{F}_{\nu \sigma}{ }^b\right),
  \end{aligned}
\label{ymactions}  
\end{equation}
where $\bar{F}_{\nu \sigma}{}^a$ is the field strength corresponding to gauge field $\bar{A}_\nu{}^a$. With these two actions, the structure of DFT action takes the form,
\begin{equation}
    S_{DFT} \equiv S_{YM1} \otimes S_{YM2}.
\end{equation}
Firstly, I consider the kinetic part of the DFT action, which corresponds to the double copy of quadratic parts of Yang-Mills actions. Passing to the momentum space, the quadratic parts of the Yang-Mills actions read,
\begin{equation}
    \begin{aligned}
        S^{(2)}_{\mathrm{YM1}} &= -\frac{1}{2} \int_k \kappa_{a b} k^2 \Pi^{\mu \rho}(k) \mathcal{A}_\mu{ }^a(-k) \mathcal{A}_\rho{ }^b(k),\\
        S^{(2)}_{\mathrm{YM2}} &= -\frac{1}{2} \int_{\bar{k}} \bar{\kappa}_{a b} \bar{k}^2 \bar{\Pi}^{\nu \sigma}(\bar{k}) \bar{\mathcal{A}}_\nu{ }^a(-\bar{k}) \bar{\mathcal{A}}_\sigma{ }^b(\bar{k}).
    \end{aligned}
\end{equation}
At this point, I follow a different route from the usual literature on the double copy formulation of DFT following \cite{Bern2010}. In \cite{Bern2010}, the authors start with the Yang-Mills action, whose quadratic part is given in Lorenz gauge by,
\begin{equation}
    \mathcal{S}^{(2)}_{Y M} = \frac{1}{2} \int d^4 k_1 d^4 k_2 \delta^4\left(k_1+k_2\right) k_2^2\left[A^\mu\left(k_1\right) A_\mu\left(k_2\right)\right].
\end{equation}
Then they write the double copy action schematically as,
\begin{equation}
\begin{aligned}
    \mathcal{S}_{DC}^{(2)} \sim \frac{1}{4} &\int d^4 k_1 d^4 k_2 \delta^4\left(k_1+k_2\right) k_2^2 \\
    &\times \left(A^\mu\left(k_1\right)A_\mu\left(k_2\right) \tilde{A}^\sigma\left(k_1\right) \tilde{A}_\sigma(k_2)\right),
\end{aligned}
\end{equation}
and obtain the gravity action by implementing the map,
\begin{equation}
    A_\mu(k) \tilde{A}_\nu(k) \rightarrow h_{\mu \nu}(k).
\end{equation}
For the cubic terms, the process is similar but now one should also be careful about the color indices. In \cite{Bern2010}, the authors encode the information of the structure constants in the anti-symmetrization and cyclicity of the interaction terms, and then turn off the color indices. For the construction of DFT action as double copy of Yang-Mills, I follow a similar route. \\
Starting with the actions (\ref{ymactions}), and using the map 
\begin{equation}
    \kappa_{ab} \rightarrow \frac{1}{2} \bar{\Pi}^{\nu \sigma}(\bar{k}),
\label{kappaaa}
\end{equation}
the double copy action at quadratic order, in momentum space, can be written schematically as,
\begin{equation}
    S_{\mathrm{DC}}^{(2)}=-\frac{1}{4} \int_{k,\bar{k}} k^2 \bar{\Pi}^{\nu \sigma} \Pi^{\mu \rho} \mathcal{A}_\mu(-k) \mathcal{A}_\rho(k)\bar{\mathcal{A}}_\nu(-\bar{k}) \bar{\mathcal{A}}_\sigma(\bar{k}),
\label{alternativedc}
\end{equation}
where I turned off the color indices by encoding the color information using (\ref{kappaaa}). Now, I define the double copy map for the fields. For the usual Yang-Mills action around Minkowski spacetime, the map can be written as
\begin{equation}
    \mathcal{A}_\mu(k) \bar{\mathcal{A}}_\nu(\bar{k}) \rightarrow e_{\mu \nu}(k,\bar{k}),
\label{usualmap}
\end{equation}
and inserting (\ref{usualmap}) into (\ref{alternativedc}) leads to the quadratic part of usual DFT action on Minkowski spacetime. To obtain DFT action on an arbitrary constant DFT background, I consider Yang-Mills action around a nontrivial background gauge field, namely,
\begin{equation}
    \mathcal{A}_\mu(k) = A_\mu + a_\mu(k),
\label{gaugefield}
\end{equation}
where $A_\mu$ is the background gauge field and $a_\mu$ is the perturbation part. Inserting (\ref{gaugefield}) into (\ref{alternativedc}) gives,
\begin{equation}
    \begin{aligned}
        S_{\mathrm{DC}}^{(2)}=-\frac{1}{4} \int_{k,\bar{k}} k^2 \bar{\Pi}^{\nu \sigma} \Pi^{\mu \rho}&(A_\mu + a_\mu(-k))(\bar{A}_\nu + \bar{a}_\nu(-\bar{k})) \\
        &\times (A_\rho + a_\rho(k))(A_\sigma + a_\sigma(\bar{k})).
    \end{aligned}
\label{alternativedc2}
\end{equation}
Now, one needs to determine double copy maps for the fields. A consistent choice for the perturbation part is
\begin{equation}
    a_\mu(-k)\bar{a}_\nu(-\bar{k}) \rightarrow e_{\mu \nu}(-k,-\bar{k}).
\label{pertdc}
\end{equation}
Similarly for the multiplication of two background gauge fields one can choose,
\begin{equation}
    A_\mu\bar{A}_\nu \rightarrow \tilde{E}_{\mu \nu},
\label{backdc}
\end{equation}
where,
\begin{equation}
    \tilde{E}_{\mu \nu} \equiv E_{\mu \nu}-\eta_{\mu \nu}.
\end{equation}
For the multiplication of background gauge field and perturbation part of the gauge field, a double copy map schematically can be given as
\begin{equation}
    A_\mu\bar{a}_\nu(-\bar{k}) \rightarrow C\tilde{E}_\mu{}^\alpha e_{\alpha \nu} \quad \text{and} \quad a_\mu(k)\bar{A}_{\nu} \rightarrow C\tilde{E}^\alpha{}_\nu e_{\mu \alpha},
\label{newmap}
\end{equation}
where $C$ is a constant which will be determined. Inserting (\ref{pertdc}),(\ref{backdc}) and (\ref{newmap}) into (\ref{alternativedc}) gives
\begin{equation}
     S_{\mathrm{DC}}^{(2)}=-\frac{1}{4} \int_{k,\bar{k}} k^2 \bar{\Pi}^{\nu \sigma}(\bar{k}) \Pi^{\mu \rho}(k) e'_{\mu \nu}(-k,-\bar{k})e'_{\rho \sigma}(k, \bar{k}),
\label{primedft}
\end{equation}
where
\begin{equation}
    e'_{\mu \nu} \equiv e_{\mu \nu} + \tilde{E}_{\mu \nu} +C (\tilde{E}_\mu{}^\alpha e_{\alpha \nu} + \tilde{E}^\alpha{}_\nu e_{\mu \alpha}).
\label{eprime}
\end{equation}
The action (\ref{primedft}) represents the quadratic part of the DFT action in Minkowski spacetime for the primed field $e'_{\mu \nu}$. The question is if this action is equal to the DFT action for $e_{\mu \nu}$ around a constant background $E_{\mu \nu}$ for some choice of constant $C$ in (\ref{newmap}), namely,
\begin{equation}
    S^{(2)}(E_{\mu \nu}, e_{\mu \nu}) \overset{?}{=} S^{(2)}(\eta_{\mu \nu}, e'_{\mu \nu}).
\label{eta=E1}
\end{equation}
An answer to this question is already given in \cite{backindepdft}, where the authors constructed a background independent form of DFT action. In \cite{backindepdft}, the authors showed that,
\begin{equation}
    S\left[E_{i j}-\chi_{i j}, e_{i j}+\chi_{i j}-f_{i j}(\chi, e)\right]=S\left[E_{i j}, e_{i j}\right]+O\left(e^3\right),
\label{backesitlik}
\end{equation}
where $\chi_{i j}$ is an infinitesimal, constant part of the field $e_{ij}$ and,
\begin{equation}
    f_{i j}(\chi, e)=\frac{1}{2}\left(\chi_i{ }^k e_{k j}+\chi^k{ }_j e_{i k}\right)+O\left(e^2\right).
\end{equation}
Choosing $\chi_{ij}=E_{ij}-n_{ij} \equiv \tilde{E}_{ij}$, one has
\begin{equation}
    f_{i j}(\chi, e)=\frac{1}{2}\left(\tilde{E}_i{ }^k e_{k j}+\tilde{E}^k{ }_j e_{i k}\right)+O\left(e^2\right),
\end{equation}
and (\ref{backesitlik}) becomes
\begin{equation}
    S\left[\eta_{i j}, e'_{i j}\right]=S\left[E_{i j}, e_{i j}\right]+O\left(e^3\right)
\label{eta=E}
\end{equation}
where 
\begin{equation}
    e'_{i j} \equiv e_{ij} + \tilde{E}_{ij}-\frac{1}{2}\left(\tilde{E}_i{ }^k e_{k j}+\tilde{E}^k{ }_j e_{i k} \right).
\label{primefield}
\end{equation}
One can see that (\ref{eta=E}) implies $C=-\frac{1}{2}$ in (\ref{eprime}) so that (\ref{eta=E1}) is satisfied. \\

Now, I continue with the cubic order. Considering the cubic part of the actions (\ref{ymactions}), the double copy action at cubic order, in momentum space, can be written as,
\begin{equation}
    \begin{aligned}
S_{\mathrm{DC}}^{(3)}= & -\frac{i g_{\mathrm{YM}}}{6(2 \pi)^{D / 2}} \int_{K_1, K_2, K_3} \delta\left(K_1+K_2+K_3\right) \\
& \times \bar{\Pi}^{\sigma \tau \kappa} \Pi^{\mu \nu \rho} \mathcal{A}_{1 \mu} \mathcal{A}_{2 \nu} \mathcal{A}_{3 \rho}\bar{\mathcal{A}}_{1 \sigma} \bar{\mathcal{A}}_{2 \tau} \bar{\mathcal{A}}_{3 \kappa}.
\end{aligned}
\end{equation}
Again, defining the gauge fields as (\ref{gaugefield}), the cubic part of the double copy action becomes
\begin{equation}
    \begin{aligned}
S_{\mathrm{DC}}^{(3)}= & \frac{1}{48(2 \pi)^{D / 2}} \int d K_1 d K_2 d K_3 \delta\left(K_1+K_2+K_3\right) \\
& \times \bar{\Pi}^{\sigma \tau \kappa}(\bar{k}_1, \bar{k}_2, \bar{k}_3) \Pi^{\mu \nu \rho}\left(k_1, k_2, k_3\right) \\
&\times(A_{\mu1}+a_{\mu1})(A_{\nu2}+a_{\nu2})(A_{\rho3}+a_{\rho3})\\
&\times(\bar{A}_{\sigma1}+\bar{a}_{\sigma1})(\bar{A}_{\tau2}+\bar{a}_{\tau2})(\bar{A}_{\kappa3}+\bar{a}_{\kappa3}),
\end{aligned}
\label{53}
\end{equation}
where the color indices are turned off and their information is encoded in the projectors. Using the maps
\begin{equation}
    \begin{aligned}
        &A_{\mu1}\bar{A}_{\sigma1} \rightarrow E_{\mu \sigma}, \\
        &a_{\mu1}\bar{a}_{\sigma1} \rightarrow e_{\mu \sigma}(k_1, \bar{k}_1), \\
        &A_{\mu1}\bar{a}_{\sigma1} \rightarrow -\frac{1}{2}E_{\mu}{}^\alpha e_{\alpha \sigma}(k_1,\bar{k}_1), \\
        &a_{\mu1}\bar{A}_{\sigma1} \rightarrow -\frac{1}{2} E^\alpha{}_\sigma e_{\mu \alpha},
    \end{aligned}
\end{equation}
the action (\ref{53}) becomes,
\begin{equation}
    \begin{aligned}
S_{\mathrm{DC}}^{(3)}= & \frac{1}{48(2 \pi)^{D / 2}} \int d K_1 d K_2 d K_3 \delta\left(K_1+K_2+K_3\right) \\
& \times \bar{\Pi}^{\sigma \tau \kappa}(\bar{k}_1, \bar{k}_2, \bar{k}_3) \Pi^{\mu \nu \rho}\left(k_1, k_2, k_3\right)e'_{1\mu \sigma}e'_{2\nu \tau}e'_{3\rho \kappa},
\end{aligned}
\label{primecubic}
\end{equation}
where prime fields defined as in (\ref{eprime}) with $C=-\frac{1}{2}$. Then, (\ref{primecubic}) defines the cubic part of the DFT action for prime field. The result of \cite{backindepdft}, namely (\ref{eta=E}), guarantees that this action is equal to the cubic part of the DFT action for $e_{ij}$, on Minkowski background. Thus, one can conclude that starting with Yang-Mills actions (\ref{ymactions}) with gauge fields (\ref{gaugefield}), and using the maps (\ref{pertdc}), (\ref{backdc}), (\ref{newmap}) with $C=-\frac{1}{2}$ yields DFT action around arbitrary constant background $E_{ij}$, up to third order.
\section{Discussion and Conclusion}
Double copy is a very powerful tool to calculate gravity amplitudes by using Yang-Mills amplitudes. In recent years, there are also many results which show the power of double copy idea for classical solutions. At the level of actions, the literature is much more limited. A promising result in that sense is the formulation of perturbative DFT action up to third order by double copy procedure. In \cite{jaram}, the authors construct DFT action on Minkowski background by using double copy of Yang-Mills theory. In this work, I offer a possible way to extend this construction to the arbitrary constant DFT background. \\ 

Inspired by the type-A double copy idea of \cite{abbas}, in section 3, I start with a Yang-Mills action which is constructed around a nontrivial background gauge field, on Minkowski background. The results of this section show that a naive application of double copy procedure on this Yang-Mills action does not yield the DFT action on an arbitrary constant DFT background. \\

In section 4, I make a slight change on how I think about double copy procedure in DFT context. The motivation of this change comes from the construction of double copy Lagrangian in \cite{Bern2010}. It is important to explain this change in here. In the usual literature on double copy construction of DFT, one maps one gauge field to the one DFT field, by mapping the color indices to the second spacetime indices. In section 4 of this paper, the product of two gauge fields is mapped to one DFT field, and the color indices are turned off after their information is encoded to the kinematic factors. This can also be interpreted as a slight notation change. However, the results of Section 4 imply that this way of thinking double copy can give us a more natural way to construct DFT action on arbitrary constant DFT background by double copy. \\

In conclusion, DFT action on an arbitrary constant DFT background can be written as double copy of two Yang-Mills fields, which are constructed around nontrivial background gauge fields, on Minkowski space. In other words, type-A double copy idea can be applied to the construction of DFT action by double copy. This construction of DFT Lagrangian on an arbitrary background by a double copy procedure can have important implications on the double copy formalism at the Lagrangian level. Noting the relation of DFT with closed massless sector of string theory and supergravity, this construction can also have further implications for these theories.
\begin{acknowledgments}
I would like to express my deepest gratitude to Özgür
Sarıoğlu for the helpful discussions, and a careful reading of this manuscript. I
am also grateful to Eric Lescano for his support, insightful comments and valuable suggestions.
\end{acknowledgments}

\bibliography{apssamp}

\end{document}